\documentclass[twocolumn, referee]{raa}   

\usepackage{graphicx,times}             
\usepackage{natbib}
\usepackage{amssymb,amsmath}
\usepackage{multirow}
\bibpunct{(}{)}{;}{a}{}{,}

\usepackage[pagebackref=true]{hyperref}
\usepackage{authblk}
\usepackage{lineno}

\newcommand{\du}[1]{{\color{black} #1}}

\begin{document}
\twocolumn
  \title{
  DeepAP: Deep Learning-based Aperture Photometry Feasibility Assessment and Aperture Size Prediction
}

   \volnopage{Vol.0 (20xx) No.0, 000--000}      
   \setcounter{page}{1}          

   \author{Zheng-Jun Du 
      \inst{1,2} 
   \and Qing-Quan Li 
      \inst{1,2} \thanks{First and second authors contribute equally to this work.}
    \and Yi-Cheng Rui$^\dagger$
      \inst{3}
   \and Yu-Li Liu
      \inst{1,2}
    \and Yu-Ting Wu
      \inst{1,2}
   \and Dong Li
      \inst{1,2}
   \and Bing-Feng Seng
      \inst{1,2}
   \and Yi-Fan Xuan
      \inst{3}
   \and Fa-Bo Feng
      \inst{3}
   }

   \institute{School of Computer Technology and Application, Qinghai University, Xining 810016, China;\\
    \and
        Qinghai Provincial Key Laboratory of Media Integration Technology and Communication, Xining, 810016, China; \\
    \and
        State Key Laboratory of Dark Matter Physics, Tsung-Dao Lee Institute \& School of Physics and Astronomy, Shanghai Jiao Tong University, Shanghai 201210, China; {\it ruiyicheng@sjtu.edu.cn}\\
\vs\no
   {\small Received 2025 May 5; accepted 2025 July 28}}

\abstract{
Aperture photometry is a fundamental technique widely used to obtain high-precision light curves in optical survey projects like Tianyu. However, its effectiveness is limited in crowded fields, and the choice of aperture size critically impacts photometric precision. To address these challenges, we propose DeepAP, an efficient and accurate two-stage deep learning framework for aperture photometry. Specifically, for a given source, we first train a Vision Transformer (ViT) model to assess its feasibility of aperture photometry. We then train the Residual Neural Network (ResNet) to predict its optimal aperture size. For aperture photometry feasibility assessment, the ViT model yields an ROC AUC value of 0.96, and achieves a precision of 0.974, a recall of 0.930, and an F1 score of 0.952 on the test set. For aperture size prediction, the ResNet model effectively mitigates biases inherent in classical growth curve methods by adaptively selecting apertures appropriate for sources of varying brightness, thereby enhancing the signal-to-noise ratio (SNR) across a wide range of targets. Meanwhile, some samples in the test set have a higher SNR than those obtained by exhaustive aperture size enumeration because of the finer granularity of aperture size estimation. By integrating ResNet with the ViT network, the DeepAP framework achieves a median total processing time of 18 milliseconds for a batch of 10 images, representing a speed-up of approximately $5.9\times 10^4$ times compared to exhaustive aperture size enumeration. This work paves the way for the automatic application of aperture photometry in future high-precision surveys such as Tianyu and LSST. The source code and model are available at \url{https://github.com/ruiyicheng/DeepAP}.
}

\authorrunning{Zheng-Jun Du, Qing-Quan Li, Yi-Cheng Rui, et al. } 
 
\titlerunning{DeepAP: Efficient and Accurate Aperture Photometry with Neural Networks }  

\maketitle


%
%

\section{Introduction}     
\label{sec:intro}

With the rapid advancement of observational technology, modern robotic photometric surveys are transforming our understanding of the time-domain universe. One of the leading facilities currently in operation is the Zwicky Transient Facility (ZTF) \citep{Bellm19}, which employs a wide-field camera on the Samuel Oschin Telescope at Palomar Observatory to scan the northern sky at the optimal cadence for supernova survey. Looking to the near future, the Vera C. Rubin Observatory Legacy Survey of Space and Time (LSST) \citep{ivezi19} is poised to revolutionize the field further.  Similarly, the upcoming Tianyu Telescope \citep{Feng24}, a one-meter robotic telescope under development in Lenghu, Qinghai, China, is designed to detect transiting planets and transient events. These surveys, both current and forthcoming, share a common need for accurate and efficient photometric measurements, which are foundational to time-domain astrophysics and the characterization of dynamic celestial phenomena.

High-precision photometry of point sources, such as stars and asteroids, is a critical step in these surveys. For example, ZTF adopts point spread function (PSF) photometry as its default technique \citep{Masci19}, which is suitable for measuring stellar fluxes in crowded fields. However, the accuracy of PSF photometry is sensitive to both the detector's sampling rate and atmospheric seeing conditions. Errors in PSF modeling can introduce biases in the resulting photometry \citep{Howell00}.

Aperture photometry is used to achieve higher precision, especially in exoplanet detection missions like Kepler\citep{Borucki10}, the Transiting Exoplanet Survey Satellite (TESS) \citep{Ricker14}, the Wide Angle Search for Planets (WASP) \citep{Pollacco06}, and the Next Generation Transit Survey (NGTS) \citep{Wheatley18}. For space-based missions such as Kepler and TESS, fixed-pixel apertures are used due to the spacecraft's excellent pointing stability \citep{Morris20}. These apertures are selected based on the pixel-wise SNR tests. However, in ground-based surveys, the need to include two additional apertures for accurate sky background subtraction makes this approach computationally expensive and operationally challenging. Furthermore, variations in the PSF, caused by changing atmospheric conditions, telescope optics, and tracking issues, complicate the task of determining the optimal aperture shape. As a practical solution, ground-based instruments typically adopt a fixed aperture shape, such as a circle or square, and adjust only the size to optimize performance.

The key parameters in aperture photometry are the sizes of three apertures: one for the target source and two for estimating the surrounding sky background. Traditionally, the inner aperture size is selected using a method called the "growth curve", which measures the SNR across a series of increasing aperture sizes and selects the size that yields the highest SNR \citep{Stetson_1990}. However, this method has several limitations. First, it only optimizes the size of the inner aperture and does not account for the background annuli. Second, it operates on a per-image basis, meaning that the optimal aperture for a single image may not generalize to the full night's data, even if that image is a stacked composite. The growth curve method maximizes the SNR for a single source in a single frame, not for the entire light curve over time. Third, it is prone to contamination from nearby stars; the algorithm may inadvertently increase the aperture size to include nearby sources, thus corrupting the target signal. In some cases, it even fails to find an optimal aperture within the tested range. In this case, aperture photometry is not feasible because the vicinity is too crowded. These limitations highlight two key challenges in aperture photometry: (1) determining whether a target is feasible for aperture photometry, and (2) if so, determining the optimal sizes for the apertures.

Recent advances in computer vision, particularly CNNs and ViTs, provide powerful tools for astronomical image analysis. CNNs excel at extracting hierarchical local features (e.g., edges, gradients) and preserving spatial relationships, making them ideal for regression and classification tasks. For example, in denoising, \citep{liu2025astronomical} proposed a self-supervised TDR method to reduce solar magnetogram noise from 8 G to 2 G while preserving faint galaxy structures in Hubble images. \citep{elhakiem2021astronomical} optimized Astro U-net for multi-noise denoising via hybrid fusion, achieving superior PSNR/SSIM improvements. \citep{vojtekova2021learning} demonstrated that Astro U-net emulates doubled exposure times, boosting SNR by 1.63× for HST data. \citep{gheller2022convolutional} employed denoising autoencoders to detect faint radio cosmic webs in Low Frequency Array (LOFAR) observations through cosmological simulations. 
\du{
For galaxy morphology classification, \citet{dieleman2015rotation} pioneered rotation-invariant CNNs for Galaxy Zoo data, achieving $>$99\% accuracy on high-agreement samples. Large-scale morphology catalogs were enabled by \citet{gravet2015catalog} (CANDELS, 50k galaxies) and \citet{dominguez2018improving} (SDSS, 670k galaxies), with the latter reducing misclassification to $<$1\%. To address limited labeled data, \citet{luo2025galaxy} proposed semi-supervised GC-SWGAN, achieving 75\% accuracy on Galaxy10 DECals using only 20\% labeled samples. For survey data enhancement, \citet{luo2025cross} developed Pix2WGAN, a CNN-based neural network, to transform SDSS / DECaLS images to HSC quality, improving structural visibility. \citet{sandeep2021analyzing} implemented multiple CNN architectures for simultaneous galaxy classification (92.3\% accuracy) and redshift prediction. For rare object identification, \citet{primack2018deep} detected ``blue nugget" galaxies in CANDELS, revealing their characteristic mass range (10$^{9.2–10.3}$ M$_\odot$), while \citet{davies2019using} achieved 77\% recall for gravitational lenses in Euclid-like simulations. Advanced techniques include \citet{burke2019deblending}'s Mask R-CNN for deblending (98\% galaxy precision) and \citet{jia2020detection}'s real time Faster R-CNN for wide-field surveys.
}


\du{In contrast, ViTs capture global contextual patterns through self-attention mechanisms, enhancing robustness for complex astronomical image classification. For example, \citep{donoso2023astromer} introduced ASTROMER, a transformer-based model that generates light curve representations via self-supervised pretraining, improving downstream classifiers in limited-label scenarios. Similarly, \citep{yang2024stellar} applied ViT (``stellar-ViT") to stellar classification using SDSS photometric images, achieving 83.9-86.3\% accuracy across seven spectral classes and outperforming CNNs. For galaxy morphology, ViT-based approaches show strong performance: \citep{bhavanam2024enhanced} used ViT-CNN hybrids to improve faint SDSS object classification. \citep{cao2024galaxy} proposed CvT (Convolutional Visual Transformer) for galaxy morphology classification. Compared with other five class classification models, CvT outperforms $>$98\% in terms of average accuracy, precision, recall, and F1 score. \citep{yeganehmehr2025classification} further demonstrated ViT's capability, attaining 99.85\% classification accuracy on galaxy images. Transformers also excel in time-series astronomy: \citep{allam2024paying} developed a time-series transformer for photometric transient classification, achieving competitive performance (log-loss 0.507, AUC 0.98) on LSST challenge data with minimal feature engineering.
}


\du{
To fully exploit the strengths of Vision Transformers (ViTs) in modeling global contextual information and Convolutional Neural Networks (CNNs) in efficient local feature extraction for our specific task, we propose a robust two-stage framework to overcome the limitations of classical aperture photometry. Specifically, stage 1 employs the ViT classifier \citep{dosovitskiy2020image} to assess the global feasibility of applying aperture photometry to a given source. If a source is deemed feasible, Stage 2 utilizes the ResNet-18 network \citep{he2016deep} to predict the optimal sizes for the three apertures. This integration leverages ViT's capability in global source viability assessment and CNN's efficiency in precise aperture size prediction, thereby enhancing the overall photometric performance.
}

The remainder of the paper is organized as follows. Section \ref{sec:method} presents the dataset and architecture design; Section \ref{sec:experiment} evaluates the performance of our method on real survey data and shows the comparison with other methods; and Section \ref{sec:conc} concludes with implications for future improvements.

\begin{figure*}[!ht]
\centering
\includegraphics[width=1\textwidth]{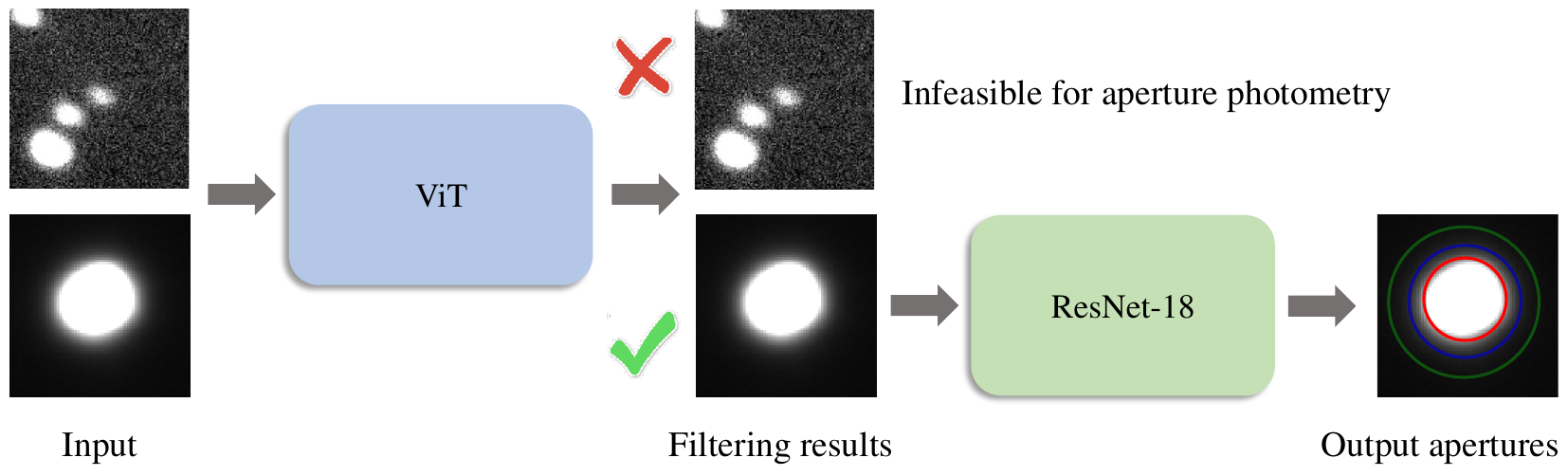}
\caption{The pipeline of our method. Our method contains two stages. In stage 1, we filter out chopped images that are infeasible for aperture photometry via a ViT network. In stage 2, we train a ResNet, to obtain the accurate apertures of a given chopped image.} 
\label{fig:pipeline}
\end{figure*}

\section{Method}     
\label{sec:method}
\subsection{Overview}
Our goal is to achieve efficient yet accurate aperture photometry. To this end, we employ a two-stage deep learning framework. First, we train a ViT model to filter out sources with severe occlusion or excessive noise contamination. Subsequently, a ResNet-18 model is trained to accurately predict the optimal aperture sizes of an individual target. The pipeline of the proposed method is illustrated in Fig.~\ref{fig:pipeline}.

\subsection{Dataset}
\label{sec:dataset}
To support the two-stage framework mentioned above, we collected 522 images from Muguang observatory (coordinate: 31$^\circ$10'07''N, 121$^\circ$ 36'21''E; altitude: 30 m) on 4 July 2024. A QHY600m \citep{alarcon23} camera is mounted on a CDK350 telescope\footnote{\url{https://planewave.com/products/cdk350/}}. The camera is set to use 1$\times$1 bin, and no overscan is used. The exposure time for each image is set to  30 seconds. The Baader CMOS L-Filter (420 – 685 nm) is used in this observation\footnote{\url{https://www.baader-planetarium.com/en/baader-uv-ir-cut-l-filter-cmos-optimized.html}}.

\textbf{Input data: }\du{ The input to our model comprises 128$\times$128 pixel sub-images centered on the centroids of detected sources in the stacked astronomical images. These sub-images are generated with bias subtraction, flat-field calibration, image stacking, and alignment using a modified version of the Tianyu pipeline\footnote{\url{https://github.com/ruiyicheng/Tianyu_pipeline}} \citep{Rui2025}. Among the 522 images, 506 high-quality images are utilized in the subsequent analysis. The source parameters, including centroid coordinates, full width at half maximum (FWHM), and other relevant metrics, are then extracted from the stacked image using \textsc{sep} \citep{Barbary16}. For each detected source, a 128$\times$128 pixel subimage centered on the source’s centroid is extracted from the stacked image to serve as input to our dataset. }

\begin{figure*}[htbp]
\centering
\includegraphics[width=1\textwidth]{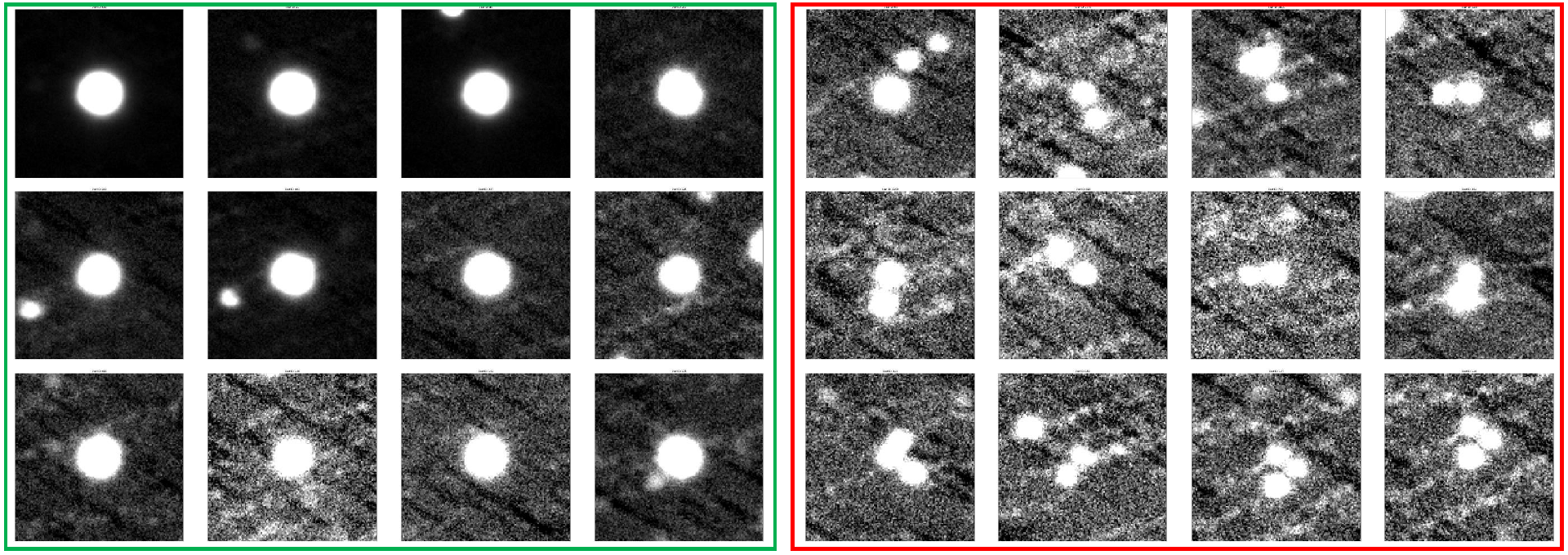}
\caption{Visualization of samples from the aperture photometry feasibility determination dataset. Images in the first four columns are labeled as 1 (feasible for aperture photometry), while the remaining columns are labeled as 0 (infeasible). The streak-like patterns in the background result from masks applied to hot pixels and do not affect the photometry. All resolved sources have been manually verified to ensure they represent genuine point sources rather than artifacts caused by these streak-like fluctuations.}
\label{fig:dataset1}
\end{figure*}

\begin{figure*}[htbp]
\centering
\includegraphics[width=1\textwidth]{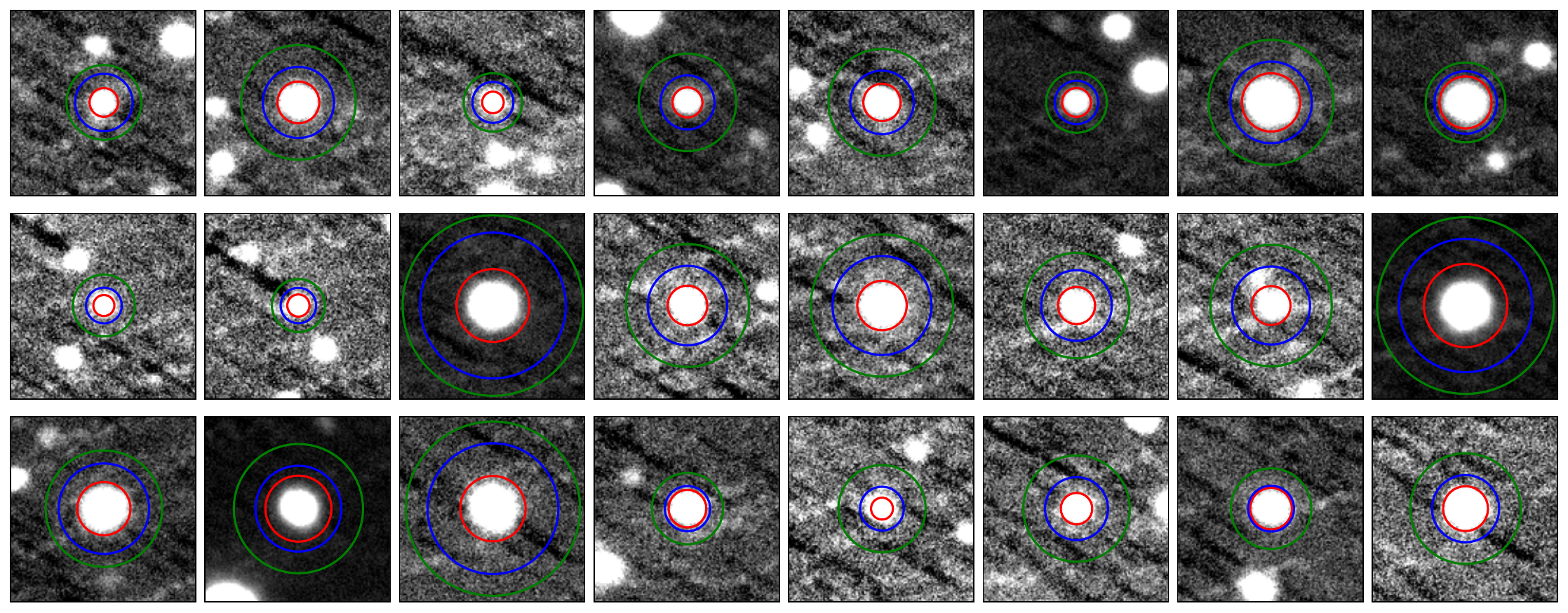}
\caption{Visualization of samples from the aperture prediction dataset (inner aperture: red, middle aperture: blue, outer aperture: green). The rings in each image represent the optimal apertures for photon counting (inner ring) and sky background estimation (outer two rings), determined through exhaustive enumeration.} 
\label{fig:dataset2}
\end{figure*}

\du{Label construction: To construct the training labels, for each source in the input data, we employ brute-force enumeration to find its optimal aperture, and then assess its feasibility for aperture photometry based on apertures of surrounding sources.
}
To optimize the aperture sizes for each object, we systematically vary the radii and select the combination that yields the light curve with the highest SNR. Specifically, for the $i$-th source, the radius of the inner aperture, $r_{\text{inn},i}$, is varied from 1 FWHM to 5 FWHM in increments of 0.05 FWHM. The radius of the middle aperture (the inner boundary of the sky background annulus), $r_{\text{mid},i}$, is varied from $1.1r_{\text{inn},i}$ to $2r_{\text{inn},i}$, in steps of $0.1r_{\text{inn},i}$. The outer aperture radius (the outer boundary for sky background estimation), $r_{\text{out},i}$, is varied from $r_{\text{mid},i} + 3$ to $r_{\text{mid},i} + 15$, in steps of 3 pixels. For each set of radii $(r_{\text{inn},i}, r_{\text{mid},i}, r_{\text{out},i})$, the corresponding light curve is extracted using \textsc{sep}. The SNR of each light curve is defined as \begin{equation} \text{SNR} = \frac{\mu_i}{\sigma_i}, \label{eq:snr} \end{equation} where $\mu_i$ and $\sigma_i$ are the weighted mean and weighted standard deviation of the light curve for the $i$-th object, respectively. The weight of each flux measurement in the light curve is assigned to be inversely proportional to the square of the photometric uncertainty for a single exposure, which are also provided by \textsc{sep}. For each object in the training set, we enumerate $(r_{\text{inn},i}, r_{\text{mid},i}, r_{\text{out},i})$ and obtain the corresponding SNR of the light curve.

To create the data set for aperture photometry feasibility classification, each image is labeled with a binary label: 0 indicates that the image is infeasible for aperture photometry due to overcrowding, while 1 indicates feasibility. Specifically, the image of the $i$-th star is labeled as 1 if and only if 
there exists a set of $(r_{\text{inn},i}, r_{\text{mid},i}, r_{\text{out},i})$ such that for all $j\ne i$,

\begin{equation}
\begin{aligned}
& r_{ij} > r_{\text{inn},i} + 3 \text{FWHM}_j; \\
& \text{and} \quad 
\bigl( r_{ij} < r_{\text{mid},i} - 3 \text{FWHM}_j \\
& \text{or} \quad r_{ij} > r_{\text{out},i} + 3 \text{FWHM}_j \bigr),
\end{aligned}
\label{eq:cond1}
\end{equation}
where $r_{ij}$ denotes the distance between the $i$-th and $j$-th resolved sources, and $\text{FWHM}_j$ is the full width at half maximum of the $j$-th source. The image is labeled as 0 otherwise. The 3 FWHM gap is required to ensure the existence of an uncontaminated aperture for photometry when a source is marked as feasible because the flux contamination of other sources is negligible at this distance\citep{Howell00}.

\du{Applying this criterion to all 2,677 sources, 2,311 are labeled as feasible (label 1). The sources are then divided into training and testing subsets in a 7:3 ratio, resulting in 1,983 sources in the training set and 804 in the test set for the aperture photometry feasibility classification task.  Representative examples from the dataset are presented in Fig.~\ref{fig:dataset1}. To augment the training set, we apply rotational transformations at angles of 36°, 72°, 108°, ..., 324° to full-frame stacked image using \textsc{SCAMP} \citep{bertin06} and \textsc{SWARP} \citep{bertin10}. The corresponding $128\times128$ image cutouts are extracted around each source using the world coordinate system (WCS) information and the right ascension/ declination positions of the 1,983 training sources. This augmentation yields a total of 19,830 $128\times128$ training images. The same labels are assigned to the augmented images, based on the assumption of rotational symmetry.}

\du{For aperture size optimization, we use the same 2311 sources labeled as feasible in the previous step. Each image corresponding to the $i-$th source is labeled with the set of aperture radii $(r_{\text{inn},i}, r_{\text{mid},i}, r_{\text{out},i})$ that maximize the SNR, as defined in Eq.~\ref{eq:snr}, under the constraint specified in Eq.~\ref{eq:cond1} for all $j \ne i$. The dataset is then split into training and testing subsets using a 7:3 ratio, yielding 1,617 training sources and 694 test sources. Representative examples from this dataset are shown in Fig.~\ref{fig:dataset2}. As before, we apply 9-fold rotational augmentation to the training sources, resulting in 16,170 $128\times128$ images. The aperture radius labels are inherited from the corresponding unrotated sources, again under the assumption of rotational symmetry.}

\begin{figure*}[htbp]
\centering
\includegraphics[width=1\textwidth]{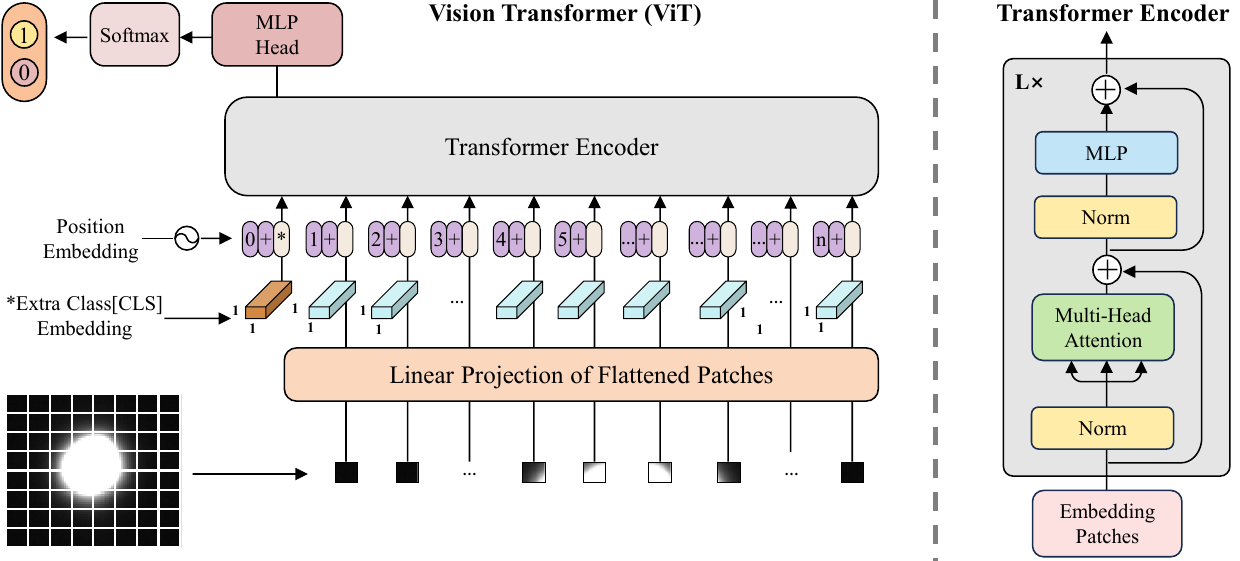}
\caption{The ViT model for aperture photometry feasibility assessment. The ViT processes an input image by first dividing it into flattened patches. Then, each patch is linearly projected into an embedding vector, combined with positional embeddings and a learnable classification token [CLS]. Next, the sequence is fed into a multi-layer Transformer encoder,  followed by a MLP block and the softmax activation function. Finally, it outputs a binary classification probability distribution.} 
\label{fig:vit}
\end{figure*}

\subsection{Aperture photometry feasibility assessment}

The first stage of our method employs a ViT network to filter out celestial objects that are infeasible for aperture photometry due to severe occlusion or excessive noise contamination. The network architecture and data flow are illustrated in Fig.~\ref{fig:vit} and are described below.

For an input clipped image $x\in \mathbb{R}^{H \times W \times C}$, where $H=W=128,C=1$  (image height, width, and channels), we first split it into $N$ patches of size $ P \times P $, with $ N = 64, P = 16 $. Each patch $ x_p^i \in \mathbb{R}^{P \times P \times C} $ is then flattened into a 1D vector and linearly projected into a higher-dimensional space using a learnable patch embedding $ E(\cdot) $. To preserve spatial information, a positional embedding $ E_{pos} \in \mathbb{R}^{N \times P^2} $ is added to the patch embeddings. Additionally, a classification token $ x_{\text{class}} $ is prepended to the sequence to aggregate global information for the final classification. The input to the Transformer Encoder is thus created as:
\begin{equation}
\small{
Z_0 = [x_{\text{class}};E(x^1_p);E(x^2_p);\cdots;E(x^N_p)] + E_{pos}.
}
\end{equation}
The input sequence $ Z_0 $ is processed by $L$ cascaded Transformer encoder layers. Each layer consists of two key components:
\begin{itemize}
\du{
    \item [1)] \textbf{Multi-head Self-Attention (MSA)}: MSA captures long-range dependencies between image patches by enabling global interactions. Its multi-head design learns diverse patterns in parallel subspaces. In implementation, each head applies independent learnable projections to transform these shared embeddings into three distinct vector sets: Queries (Q), Keys (K) and Values (V). Within each head, every query interacts with all keys to compute attention weights that dynamically blend values. This allows global context aggregation while maintaining head-specific perspective diversity. The combined head outputs are linearly fused, implemented as: \begin{equation}
        Z'_l = \text{MSA}(\text{LN}(Z_{l-1})) + Z_{l-1}.
    \end{equation}
    }
    
    \item [2)] \textbf{Feed-Forward Network (MLP)}: A two-layer perceptron with non-linearity, applied to each token independently. Similar to MSA, layer normalization and residual connections are used:
    \begin{equation}
        Z_l = \text{MLP}(\text{LN}(Z'_l)) + Z'_l.
    \end{equation}
\end{itemize}
The final output token $ Z_L^0 $ (corresponding to the classification token) is passed through a Multi-Layer Perceptron (MLP) followed by a softmax activation, producing a probability distribution $ y = [y_0, y_1] $ over the two classes:
\begin{equation}
    y = \text{Softmax}(\text{MLP}(Z_L^0)),
\end{equation}
where $ y_0 $ and $ y_1 $ represent the probabilities of the input image being infeasible (label 0) or feasible (label 1) for aperture photometry, respectively.

\begin{figure*}[!ht]
\centering
\includegraphics[width=1\textwidth]{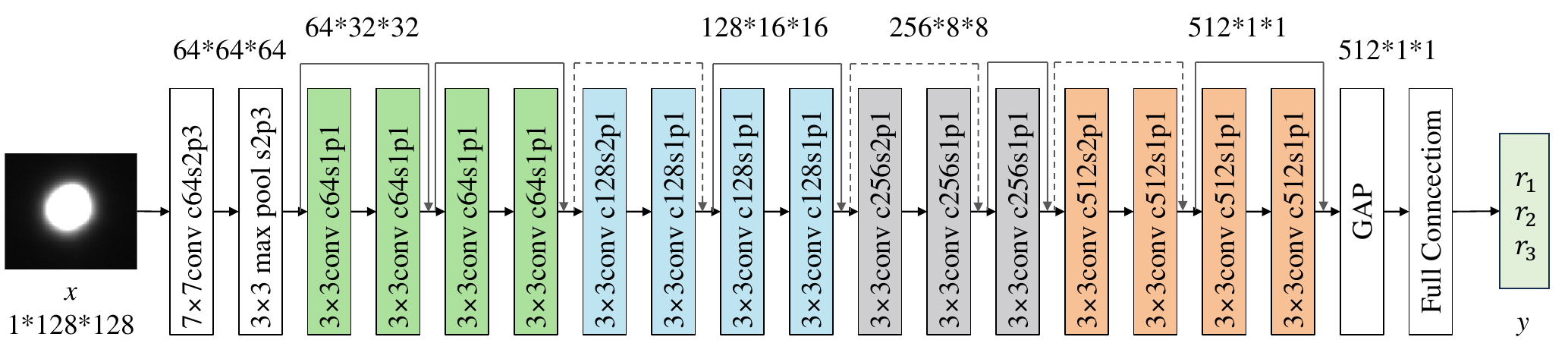}
\caption{The ResNet-18 model for optimal aperture size estimation. The numbers following ``c", ``s", and ``p" denote the channel, stride, and padding parameters of each layer, respectively. Solid arrows indicate direct forward connections, while dashed arrows represent down-sampling operations implemented with 1$\times$1 convolution layers of stride 2.} 
\label{fig:resnet}
\end{figure*}

The network is trained using Binary Cross-Entropy (BCE) loss, defined as:
\begin{equation}
    \mathcal{L}_{\text{fa}} = -p_0 \log(y_0) - p_1 \log(y_1),
\end{equation}
where $ p = [p_0, p_1] $ is the ground-truth label (one-hot encoded). This stage effectively filters out low-quality images, ensuring that only reliable candidates proceed to aperture prediction.

\subsection{Optimal aperture size estimation}
In this stage, we adopt ResNet-18, a well-proven Convolutional Neural Network in computer vision, as the backbone to predict the optimal apertures of a given source. The ResNet-18 network is illustrated in Fig.~\ref{fig:resnet}, which takes a 128$\times$128 chopped image as input and outputs three aperture values indicating the inner ($r_1$), middle ($r_2$) and outer ($r_3$) radii, respectively.

The ResNet-18 network consists of a series of convolutional layers, followed by a global average pooling (GAP) layer and a fully connected layer with three neurons. The convolutional layers are organized into eight “ResNet blocks”, each containing a set of convolutional layers with skip connections that allow the network to learn residual functions. This design enables the network to achieve deeper architectures while mitigating the problem of vanishing gradients during training.

The data flow through the network can be described as follows:

\begin{itemize}
  \item [1)] \textbf{Input}: A 128$\times$128 pixel single-channel chopped image is fed into the network.

  \item [2)] \textbf{Initial Convolution and Pooling}: The input image is first processed by a 7$\times$7 convolutional layer with 64 output channels and a stride of 2, followed by a 3$\times$3 max pooling layer with a stride of 2. This initial block reduces the spatial resolution of the image while increasing the depth.

  \item [3)] \textbf{ResNet Blocks}: The output of the initial block is then passed through seven ResNet blocks, each containing a series of 3$\times$3 convolutional layers with 64, 128, 256, or 512 output channels depending on the block’s position in the network. The skip connections within each block allow the network to learn the identity function, enabling it to learn more complex relationships between pixels in the input image.

  \item [4)] \textbf{Global Average Pooling}: After the ResNet blocks, the network applies a global average pooling layer, which reduces the spatial dimensions of the feature maps to a single value per feature map.

  \item [5)] \textbf{Fully Connected Layer}: The output of the GAP layer is then passed to a fully connected layer with three neurons. This layer outputs the three real numbers representing the sizes of the three circular apertures.
\end{itemize}
We employ the Mean Squared Error (MSE)
to measure the difference between the network’s predicted aperture sizes $\left(r_1,r_2,r_3\right)$ and the ground truth values $\left( r_{\text{inn}},r_{\text{mid}},r_{\text{out}}\right)$:
\begin{equation}
  \mathcal{L}_{\text{ap}} = \frac{1}{3} \left\|\left(r_1,r_2,r_3\right) -\left( r_{\text{inn}},r_{\text{mid}},r_{\text{out}}\right)\right\|^2 \label{eq:loss}
\end{equation}

The ResNet-18 architecture offers a robust and efficient framework for learning features from a chopped image and regressing aperture sizes. Skip connections and deep network structure enable the network to capture complex patterns and relationships in the data, leading to accurate and reliable predictions.

\section{Experiments}     
\label{sec:experiment}
All of our experiments were conducted on a computer with the following configurations: An 13th Gen Intel(R) Core(TM) i7-13700 CPU (14 cores/20 threads), 2.10 GHz, coupled with 128GB memory. In addition, the system is equipped with an NVIDIA GeForce RTX 4090 GPU that features 24GB of GDDR6X VRAM,  16384 CUDA cores, and delivers 82.6 TFLOPS of single-precision floating-point performance. 

\subsection{Training}
In our two-stage framework, the ViT model is trained for 50 epochs with a batch size of 256, while the ResNet-18 model is trained for 100 epochs with a reduced batch size of 12. Both models use the same Adam optimizer and learning rate ($0.0001$).



For the ViT model, the training loss and testing loss show a fast decline. Eventually, after 50 epochs, the training loss stabilizes around 0.10, and the testing loss stabilizes around 0.21.  The consistent downward trend and convergence of both training and testing losses indicate effective optimization without overfitting. This alignment suggests that the model successfully captures meaningful patterns in the data, achieving relatively accurate classification.

The ResNet-18 model exhibits rapid convergence in the training set, with a loss rapidly decreasing from 0.382 to 0.074 within the first 20 epochs. After 100 epochs, the training loss stabilizes at a low value of approximately 0.007. The test loss remains consistently, but higher than the training loss throughout the training process. The test loss rapidly converges after the first epoch, and after 100 epochs, the testing loss is approximately 0.240. \du{As the loss of ResNet-18 model as defined in Eq.~\ref{eq:loss}, a loss of 0.240 corresponds to an average aperture prediction error of $\sqrt{0.240}\approx0.5$ pixels. Given that the step size for enumerating the outer radius is 3 pixels, the theoretical loss for a model possessing infinite granularity, which accounts for the discretization limit, is approximately:
$\frac{1}{3}\int_{0}^{3/2} x^2 \mathrm{d}x = 0.375$.
This value reflects the expected loss when the true optimal radius lies between the enumerated values. The test loss of 0.240 is well below this bound, indicating strong generalization capabilities despite the apparent discrepancy. The observed loss is primarily attributable to the granularity of the radius enumeration, rather than to substantial overfitting. }

\subsection{Performance on test set}\label{sec:performanceontest}
\begin{figure}
    \centering
    \includegraphics[width=1\linewidth]{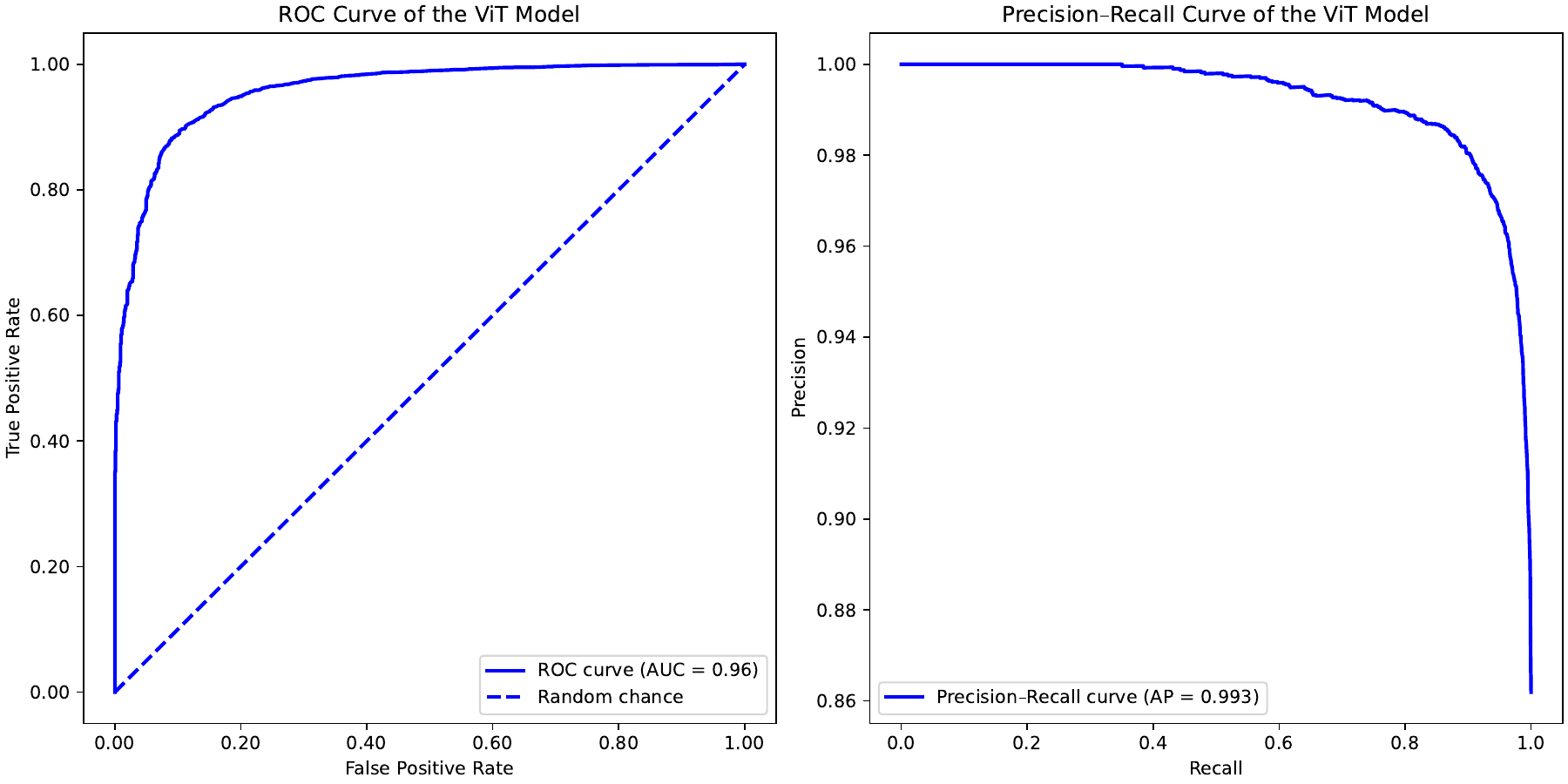}
    \caption{The ROC and Precision-Recall curves of the ViT Model.}
    \label{fig:enter-label}
\end{figure}

 \du{For the aperture photometry feasibility assessment, the confusion matrix of the ViT model is presented in Table~\ref{tab:confusion_matrix}. It achieves a precision of 0.974, a recall of 0.930, and an F1 score of 0.952 (support: 6,930). The performance of the ViT model on the test set is illustrated in Fig.~\ref{fig:enter-label}. The left panel shows the ROC curve (AUC = 0.96), while the right panel presents the precision-recall curve (AP = 0.993).
 }  The ViT model exhibited a median inference time of 4 ms for a batch size of 10 images, corresponding to an average of 0.4 ms per image.

\begin{table}[ht]
  \centering
    \begin{tabular}{|l|c|c|}
      \hline
         & Pred.  Fea.& Pred.  Infea. \\ 
      \hline
      \textbf{Actual Feasible} & TP: 6446  &  FN: 484  \\ 
      \hline
      \textbf{Actual Infeasible} & FP: 173 &  TN: 937  \\ 
      \hline
    \end{tabular}%
  
  \caption{\du{The confusion matrix of the ViT model (threshold: 0.5).}}
  \label{tab:confusion_matrix}
\end{table}


For the aperture size determination task,  we compare the performance of the ResNet-18 model with the following baseline models:
\begin{enumerate}
\item GCS (growth curve on the stacked image): This method applies the ``growth curve" approach to the stacked image, with background subtraction performed using \textsc{sep}.
\item GCF (growth curve on the first image): This method applies the ``growth curve" approach to the first calibrated image, with background subtraction performed using \textsc{sep}.
\item Exhaustive Enumeration: The optimal aperture size obtained through an exhaustive search. This method was also used to generate the ground truth labels for the training and testing datasets, as described in Section \ref{sec:dataset}.
\end{enumerate}

Optimal aperture sizes using the growth curve method are determined by varying the trial aperture radius from 1 FWHM to 5 FWHM in 0.05 FWHM steps. In the unrotated test set, comprising 694 sources, we assess the feasibility of obtaining an optimal aperture for the GCS and GCF methods. Feasibility as defined by the first condition of Eq. \ref{eq:cond1} is achieved for 658 sources with GCS and 692 sources with GCF. The ResNet-18 model is also applied to estimate aperture sizes on the same test set, yielding 600 sources satisfying the feasibility condition in Eq. \ref{eq:cond1}. To ensure a consistent comparison across GCS, GCF, and ResNet-18, we considered only the sources where a feasible aperture is obtained by all three methods. This intersection results in a common sample of 587 sources, which serve as the basis for all subsequent comparisons.

\begin{figure}[htbp]
    \centering
    \begin{subfigure}[b]{0.45\textwidth}
        \centering
        \includegraphics[width=\textwidth]{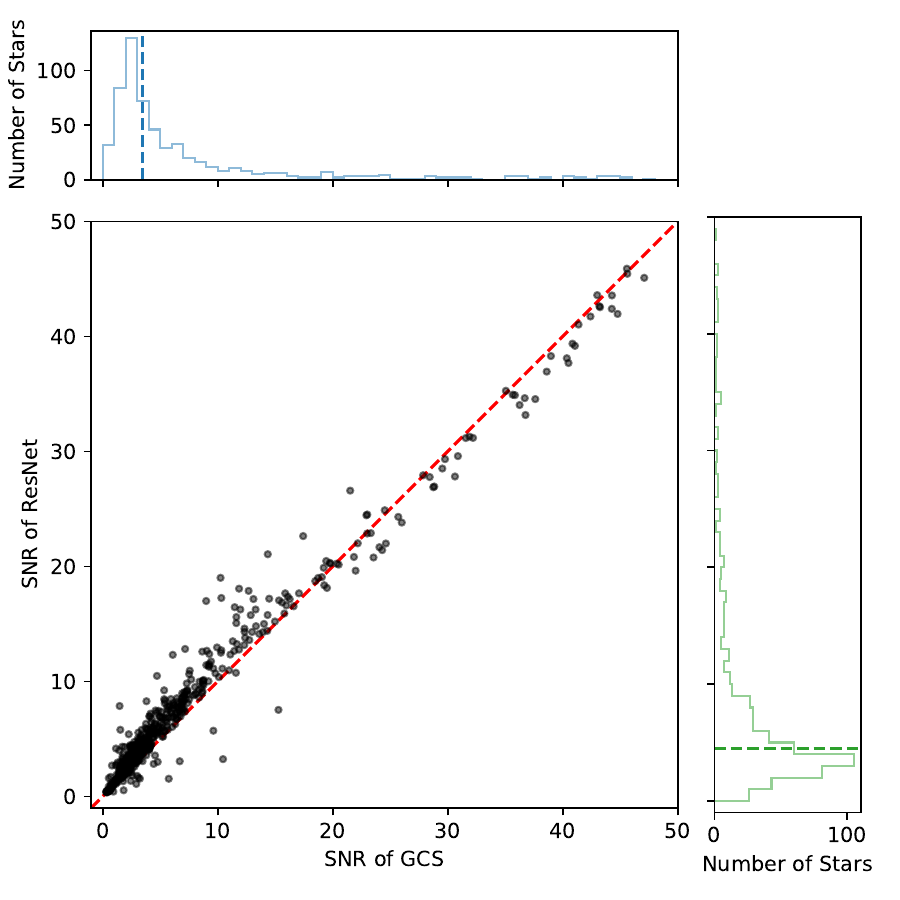}
        \vspace{-0.9cm}
        \caption{ResNet-18 vs. GCS (median SNR: 3.47)}
    \end{subfigure}
    \hfill
    \begin{subfigure}[b]{0.45\textwidth}
        \centering
        \includegraphics[width=\textwidth]{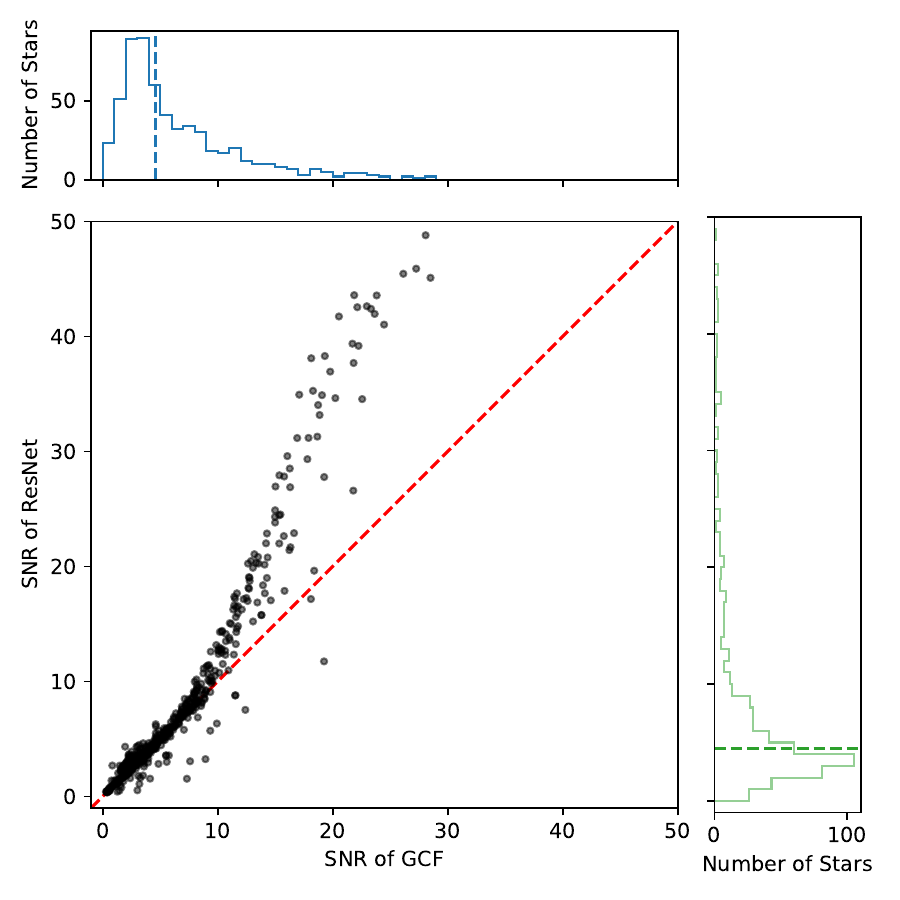}
        \vspace{-0.9cm}
        \caption{ResNet-18 vs. GCF (median SNR: 4.60)}
    \end{subfigure}
    \\[0.5em] 
    \begin{subfigure}[b]{0.45\textwidth}
        \centering
        \includegraphics[width=\textwidth]{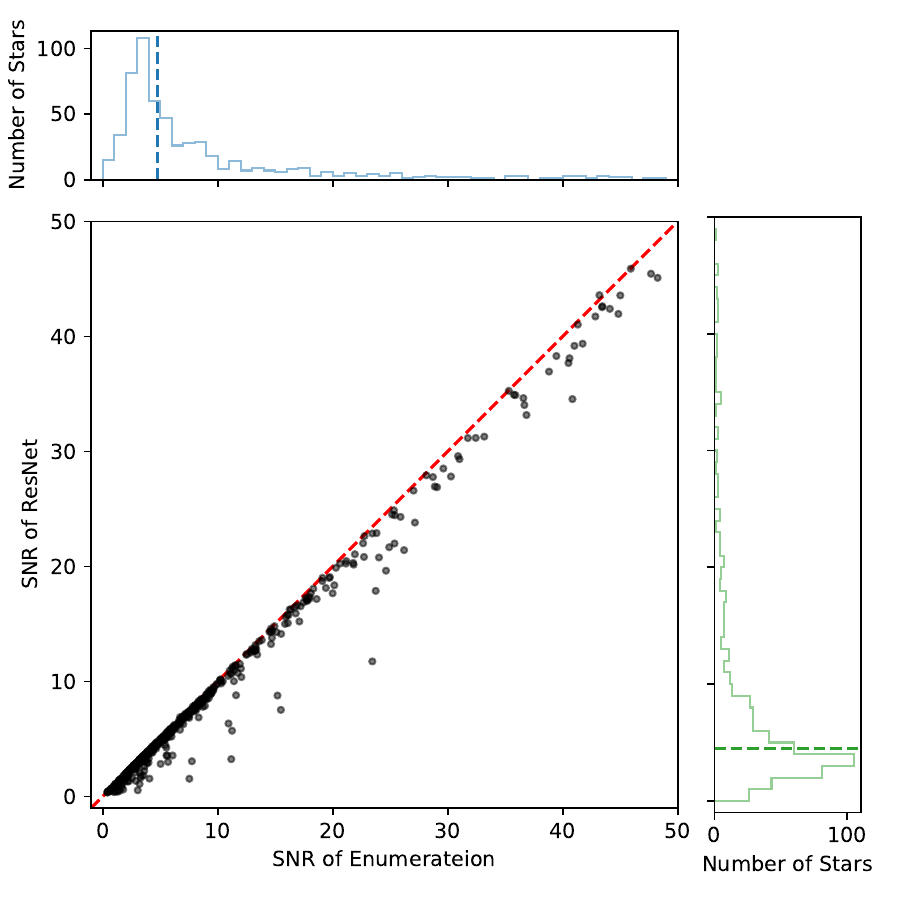}
        \vspace{-0.9cm}
        \caption{ResNet-18 vs. enumeration (median SNR: 4.79)}
    \end{subfigure}
    \caption{Comparison of the SNRs of light curves obtained with our ResNet-18 (Median SNR: 4.48), two growth curve methods, and the aperture enumeration method. The dashed lines in the histograms indicate the median SNR for each method.}
    \label{fig:comp}
\end{figure}

\begin{figure}[!htbp]
    \centering
    \includegraphics[width=0.8\linewidth]{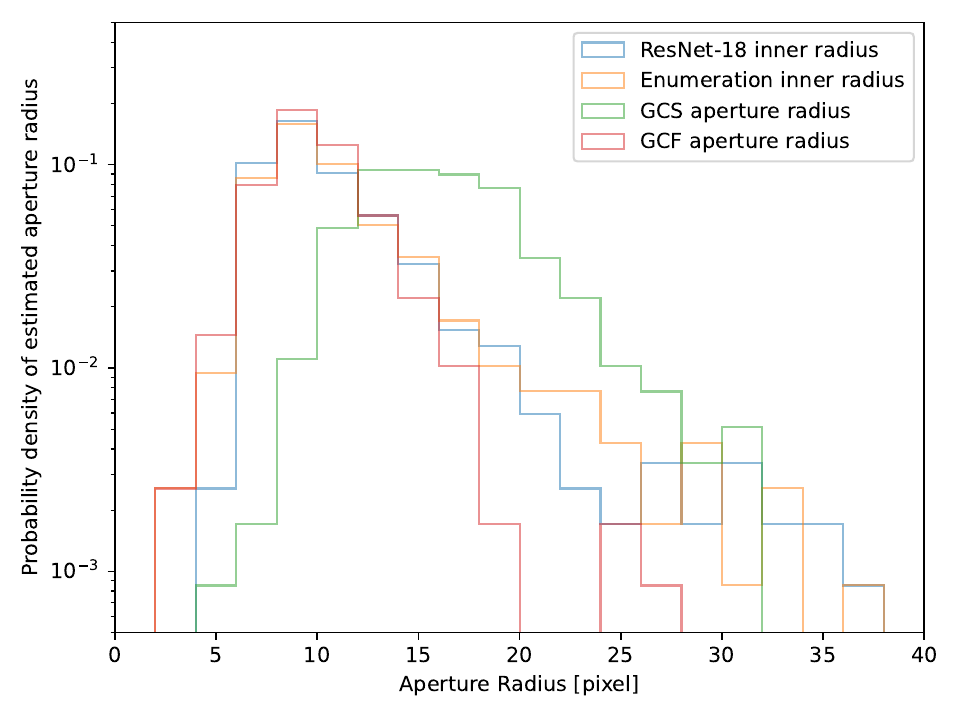}
    \caption{Distribution of the estimated inner aperture size of ResNet-18 compared with the results of aperture enumeration and growth curve. }
    \label{fig:aperture_size}
\end{figure}

\begin{figure}[!htbp]
    \centering
    \includegraphics[width=0.8\linewidth]{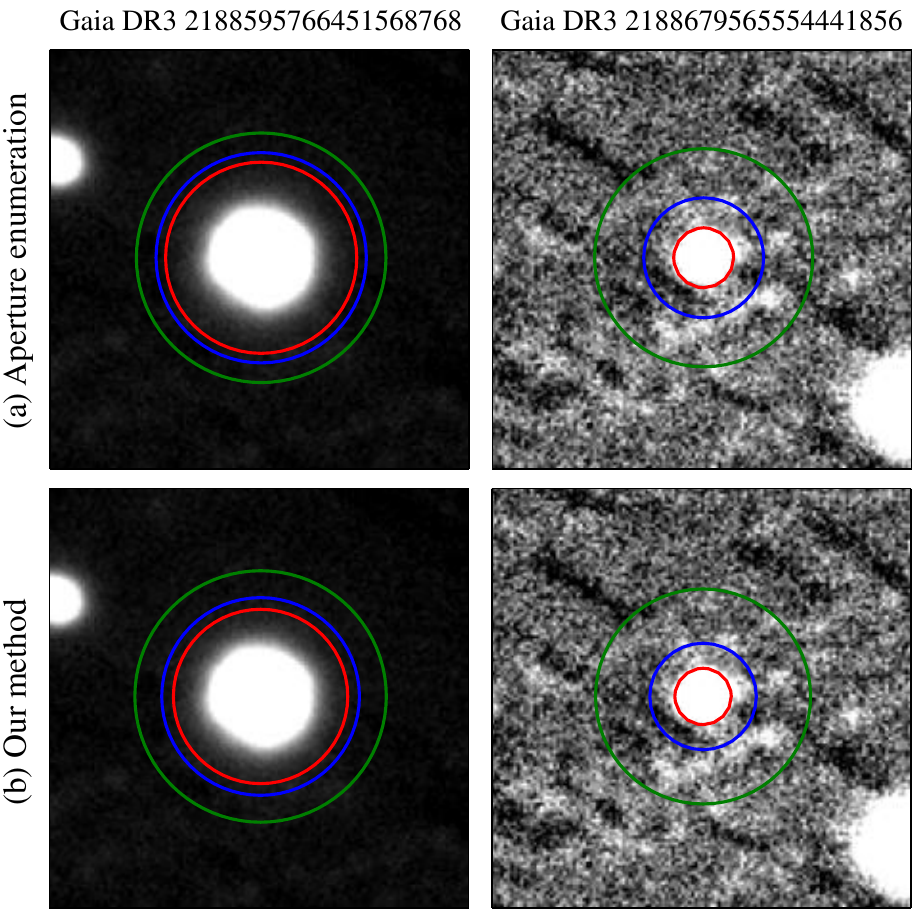}
    \caption{Comparison of the predicted aperture sizes from our method (ResNet-18) with those obtained through aperture enumeration. In both cases, the apertures predicted by our method achieve a higher SNR than those from aperture enumeration.}
    \label{fig:1459_1679_aperture_size}
\end{figure}

Fig.~\ref{fig:comp} presents a comparison of SNRs generated by different methods,
as defined in Eq.\ref{eq:snr}. The SNR obtained using estimated apertures from ResNet-18 is higher than that of the growth curve method applied to the stacked image (GCS) for 83.5\% of the sources. Meanwhile, ResNet-18 outperforms the growth curve method applied to the first calibrated image (GCF) for 66.4\% of the targets. Specifically, ResNet-18 demonstrates superior performance compared to GCS for faint targets with low SNR, and better performance than GCF for bright targets with high SNR. Fig. \ref{fig:aperture_size} shows a comparison of the inner aperture sizes selected by ResNet-18\du{, aperture size enumeration} and the growth curve methods. We observe that GCS typically selects larger apertures than GCF. This difference can be attributed to the characteristics of the images: the stacked image benefits from averaged-out sky background fluctuations, allowing larger apertures to capture more signal photons. However, this strategy is less effective for faint sources, as it also incorporates more background noise. In contrast, GCF, applied to a single, potentially noisier image, selects a smaller aperture to better match the source profile and minimize background noise. While this minimizes noise, it leads to poorer performance for bright sources by excluding some source photons. The aperture size selected by ResNet-18 adaptively behaves similarly to GCS for bright sources and GCF for faint sources. 
Meanwhile, Fig. \ref{fig:comp}(c) compares the performance of ResNet-18 with that of the exhaustive enumeration method, which serves as our training labels. Surprisingly, for 13 targets, our method outperforms the original training labels. Examples of improved SNR using ResNet-18 are shown in Fig. \ref{fig:1459_1679_aperture_size}. For instance, in the case of Gaia DR3 2188595766451568768 \du{(Gaia G-band magnitude: $G=12.2255$)}, a representative example of a bright star with a large FWHM, the ResNet-18 output exceeds the resolution of the training labels. This improvement is attributed to the finer output granularity of ResNet-18 compared to the step size used in the enumeration trials. Meanwhile, for Gaia DR3 2188679565554441856 \du{($G=17.0869$)}, the performance gain arises because the optimal outer aperture radius lies beyond the enumeration range ($r_{\rm out} - r_{\rm mid} = 16.45$), a situation frequently encountered with faint stars. The median inference time for the ResNet-18 model is 14 milliseconds when processing a batch of 10 images. \du{Meanwhile, we observe that the inner radius estimated by ResNet-18 closely matches that obtained by enumeration. The Kolmogorov–Smirnov test yields a p-value of 0.012, indicating no statistically significant difference between the distributions at the 3$\sigma$ level. 
These results indicate that ResNet-18 successfully mitigates the inherent biases of classical growth curve methods across different source brightnesses and image types. }

\du{Integrating the ResNet-18 with the ViT network, the DeepAP framework achieves a total processing time of 18 milliseconds for 10 images, representing a speed-up of approximately $5.9\times 10^4$
  times compared to exhaustive aperture size enumeration. These results highlight the strong generalizability and high speed of DeepAP. }

  \begin{figure}[h]
  \centering
  \begin{subfigure}{0.35\textwidth}
    \centering
    \includegraphics[width=\linewidth]{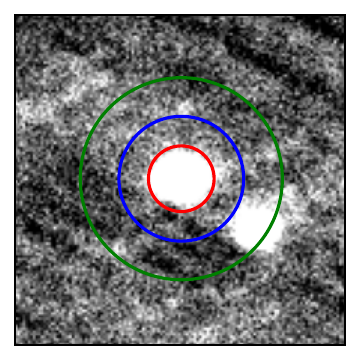}
    \caption{\du{Aperture from enumeration overlaid on stacked image}}
    \label{fig:vit582a}
  \end{subfigure}\hfill
  \begin{subfigure}{0.35\textwidth}
    \centering
    \includegraphics[width=\linewidth]{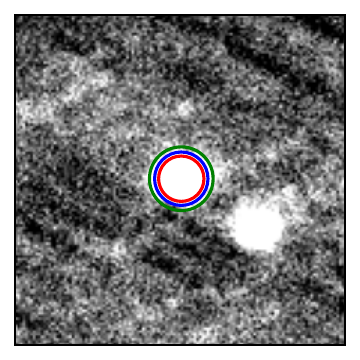}
    \caption{\du{Aperture from ResNet-18 overlaid on stacked image}}
    \label{fig:vit582b}
  \end{subfigure}
  \begin{subfigure}{0.35\textwidth}
    \centering
    \includegraphics[width=\linewidth]{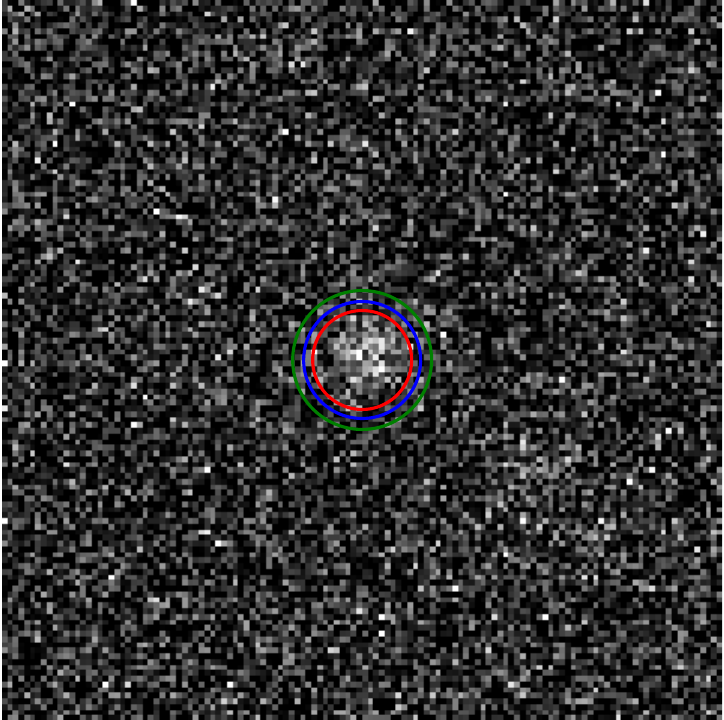}
    \caption{\du{Aperture from ResNet-18 overlaid on first calibrated image}}
    \label{fig:vit582c}
  \end{subfigure}
  \caption{\du{Qualitative comparison of aperture selection between enumeration and ResNet-18 for an infeasible case misclassified by the ViT model.}}
  \label{fig:quality-comp}
\end{figure}

\du{\subsection{Case Study: Gaia DR3 2188697333838401536}}

\du{
Gaia DR3 2188697333838401536 (right ascension: 305.537403919$^\circ$; declination: 59.563370877$^\circ$) is a relatively faint star ($G=15.9809$) within our field of view. This object serves as a false positive example for the ViT model: while its true label is 0 (infeasible), the model incorrectly classifies it as 1 (feasible), with a softmax output exceeding 0.5. Fig.~\ref{fig:quality-comp} presents the apertures selected by both the enumeration baseline and the ResNet-18 model, overlaid on the stacked image and the first calibrated image of the source.
}

\du{
Interestingly, the ResNet-18 model identifies a feasible aperture (as defined in Eq.~\ref{eq:cond1}) that successfully avoids background contamination, whereas the enumeration method fails to do so. The inner aperture radius predicted by ResNet-18 is 0.97 FWHM—slightly smaller than the lower bound used in enumeration—highlighting the limitation of the latter’s fixed search space. Although the ResNet-18 aperture may appear too small in the stacked image (Fig.~\ref{fig:vit582b}), it aligns well with the source profile in the first calibrated image (Fig.~\ref{fig:vit582c}), consistent with the analysis in Section~\ref{sec:performanceontest}. Furthermore, as shown in Appendix~\ref{app:injection}, the ResNet-18 aperture provides photometric robustness against contamination from a nearby source located at the lower right. This example underscores the strong generalization capability of DeepAP.
}

\section{Conclusion and Discussion}     
\label{sec:conc}
In this study, we introduced DeepAP, an efficient and accurate two-stage deep learning framework for aperture photometry, specifically designed to meet the demands of modern and future large-scale astronomical surveys. By combining a ViT for filtering infeasible sources for aperture photometry and a ResNet-18 network for predicting optimal aperture sizes, our method addresses the corresponding two major challenges traditionally encountered in aperture photometry: determining target suitability and selecting optimal aperture parameters.

Experimental results demonstrate that DeepAP achieves strong performance in both feasibility assessment and aperture size determination tasks. The ViT model achieves an area under the ROC curve (AUC) of 0.96, and achieves a precision of 0.974, a recall of 0.930, and an F1 score of 0.952 on the test set.
For aperture size determination, the ResNet model effectively mitigates the biases inherent in classical growth curve methods by adaptively selecting apertures appropriate for sources with different brightness, leading to enhanced SNR across a wide range of targets. By integrating ResNet with the ViT network, the DeepAP framework achieves a median total processing time of 18 milliseconds for a batch of 10 images, representing a speed-up of approximately $5.9\times 10^4$ times compared to exhaustive aperture size enumeration.

These results suggest several important implications:
\vspace{-1em}
\begin{enumerate}
    \item \textbf{Scalability and Automation:} DeepAP is capable of processing large volumes of sources rapidly and automatically, making it ideal for surveys such as LSST and Tianyu, where manual photometric tuning is infeasible.
    \item \textbf{Generalization for more sources:} By successfully applying DeepAP to targets beyond the limits of exhaustive enumeration, aperture photometry can be extended to sources traditionally considered too crowded for such techniques, thereby broadening the range of scenarios in which aperture photometry can be effectively applied. \du{We need to notice that the example as shown in Fig.~\ref{fig:quality-comp} and Appendix \ref{app:injection} does not guarantee that the ResNet-18 model can always find an feasible aperture. Meanwhile, we cannot simply assume that all the false positive case of the ViT model represents the error of labels. The ResNet-18 model does not find feasible apertures for other ViT false positive cases, which should be considered as the failure cases of this method. In real utilization, the logit threshold should be carefully chosen to control the false positive rate.}
\end{enumerate}
Despite its advantages, DeepAP still has opportunities for future improvements. Our training dataset was based on images from a single telescope-camera setup under specific observing conditions. It is worth noting that DeepAP's performance depends on the training dataset, and retraining or fine-tuning will be necessary when applying it to different telescopes or observational conditions. Transfer learning or domain adaptation techniques could also be considered. Furthermore, when performing relative flux calibration, the aperture size should be consistent within a group of stars to ensure that a comparable proportion of photons is measured for each source. In future work, we can apply more advanced numerical techniques \du{like GPU acceleration} to expand the range of aperture enumeration and improve the training set's data quality. Meanwhile, an ensemble of images should be used as input, with the model simultaneously determining both the groupings for relative photometry and the optimal aperture size for each group, making it more effective in practice.

In conclusion, DeepAP offers a practical, scalable, and high-precision method to determine the feasibility and the aperture size of aperture photometry, aligning well with the needs of next-generation time-domain astronomy project like LSST and Tianyu. Its deployment in future surveys has the potential to substantially enhance the scientific return by enabling accurate and efficient light curve extraction at scale.
\section{Acknowledgements}     
\label{sec:ack}
This work is supported by the Youth Program of the Natural Science Foundation of Qinghai Province (2023-ZJ-951Q), and Qinghai University Research Ability Enhancement Project (2025KTSQ26). 
\appendix

\section{\du{Effects of contamination of nearby variable sources} }
\label{app:injection}
\du{
In this appendix, we conduct an injection test to illustrate the effects of contamination caused by nearby variable sources and the importance of choosing an aperture to avoid it. The target shown in Fig. \ref{fig:quality-comp} is used as an example.} 

\du{In this test, we use two sets of aperture radius for the injection test:}

\du{
\begin{enumerate}
    \item The aperture obtained through enumeration yields the optimal SNR for the light curve, as shown in Fig.~\ref{fig:vit582a}. The SNR of the light curve on the image before injection, defined in Eq.~\ref{eq:snr}, is 9.89. However, this aperture is contaminated by a nearby source (Gaia DR3 2188697333838401152) located at the lower right. In this case, the aperture parameters are $(r_{\rm inn}, r_{\rm mid}, r_{\rm out}) = (12.74, 24.20, 39.20)$.
    
    \item The aperture obtained by the ResNet-18 model yields an SNR of 5.11, which is lower than that of the enumerated aperture. Here, the aperture parameters are $(r_{\rm inn}, r_{\rm mid}, r_{\rm out}) = (8.80, 10.36, 12.35)$. This aperture is smaller, and therefore collects fewer photons than the enumerated one. However, it avoids contamination from the star in the lower right.
\end{enumerate}}

\du{
The injection $\delta$ upon the pixel $(x,y)$ of the $i$-th chopped image is
\begin{align}
    \delta(\Delta {\rm mag}) 
    &= \sin\left(2\pi \frac{i}{506}\right) \cdot \left(100^{\Delta {\rm mag}/5}-1\right) \nonumber \\
    &\quad \cdot A_c \exp\left[\frac{(x-x_c)^2+(y-y_c)^2}{2 w_c^2}\right],
\end{align}
where $A_c = 11$, $x_c = 93$, $y_c = 44$, $w_c = 5$ represents the profile of the contamination star (Gaia DR3 2188697333838401152) in the lower right of Fig.~\ref{fig:quality-comp}; $\Delta{\rm mag}$ is the variational magnitude of the contamination star . The original flux of the contamination source ($G=16.7870$) is about half that of the target source ($G=15.9808$).
}

\du{
We calculate the SNR of light curve obtained by these two apertures at different $\Delta \rm mag$. The results are shown in Fig.~\ref{fig:apercompsnr}. The SNR of the light curve obtained from the enumeration aperture decrease with respect to larger variation of the contamination, while the aperture obtained by ResNet-18 is not affected by the contamination.  Meanwhile, the light curves at $\Delta\rm mag = 0$ and $\Delta \rm mag = 3$ is shown in Fig.~\ref{fig:apercomplc}. As the variability of the contaminating star increases, the flux extracted by the contaminated aperture increasingly reflects the contaminant’s signal rather than that of the target source. In cases of high contaminant variability ($\Delta \rm mag = 2.0$), the extracted light curve from the contaminated aperture even exhibits behavior opposite to the expected signal, driven primarily by fluctuations in the contaminant's flux rather than the target’s. Therefore, the aperture extracted by ResNet-18 is more robust to the contamination star variation in this case.
\begin{figure}
    \centering
    \includegraphics[width=0.9\linewidth]{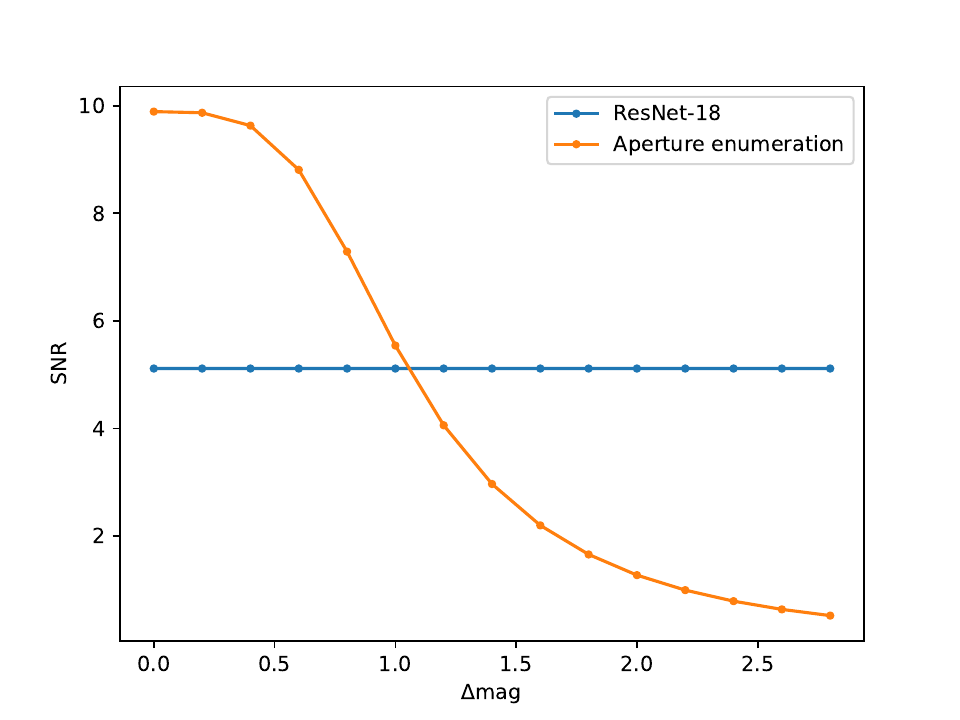}
     \vspace{-0.5cm}
    \caption{\du{SNR comparison between ResNet-18 and the enumeration aperture across different variation of the contaminating source.}}
    \label{fig:apercompsnr}
     
\end{figure}

\begin{figure}[h]
  \centering
  \begin{subfigure}{0.46\textwidth}
    \centering
    \includegraphics[width=\linewidth]{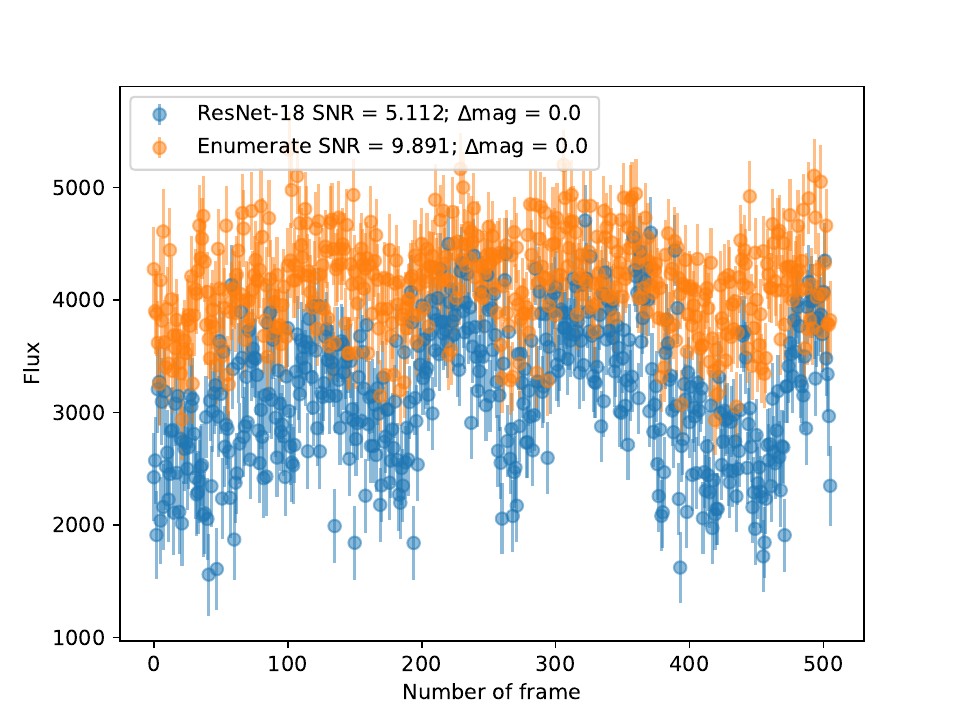}
     \vspace{-0.9cm}
  \end{subfigure}\hfill
  \begin{subfigure}{0.46\textwidth}
    \centering
    \includegraphics[width=\linewidth]{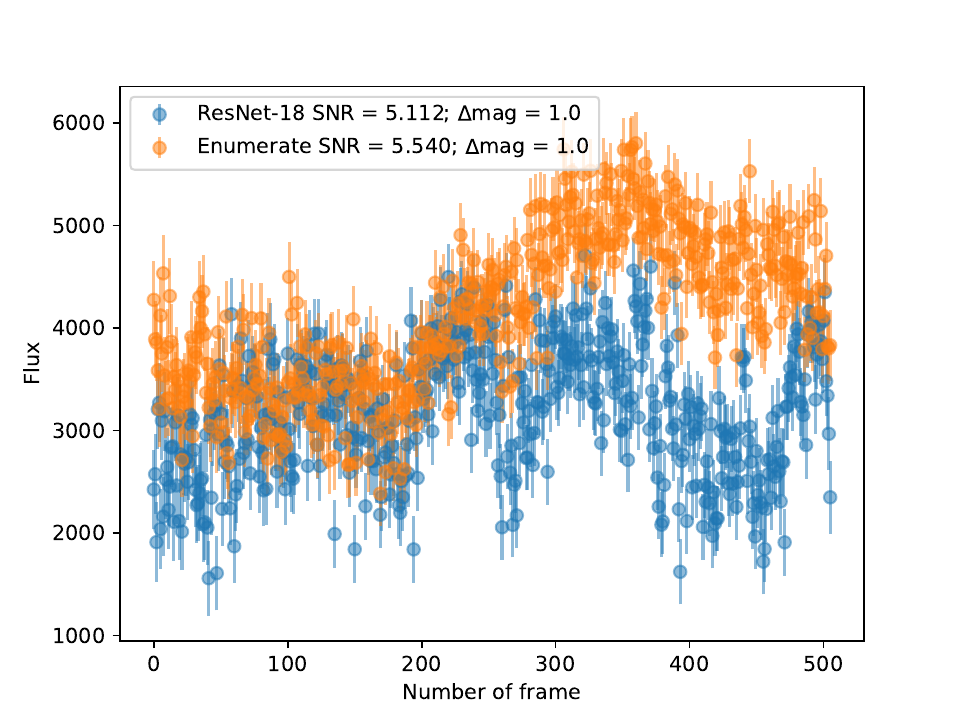}
     \vspace{-0.9cm}
  \end{subfigure}
    \begin{subfigure}{0.46\textwidth}
    \centering
    \includegraphics[width=\linewidth]{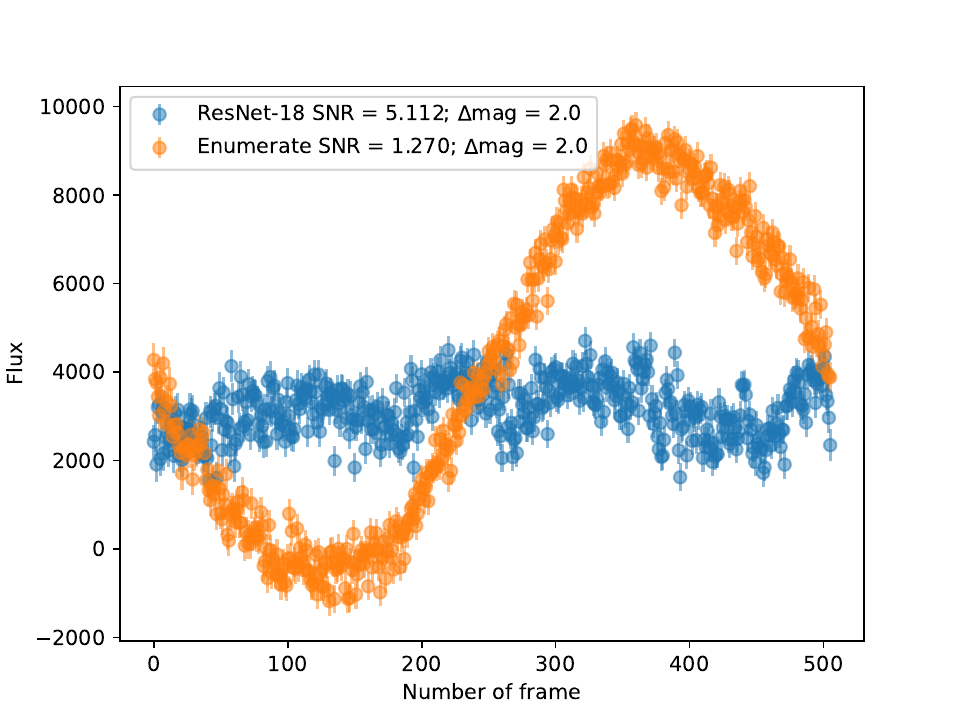}
     \vspace{-0.9cm}
  \end{subfigure}
  \caption{\du{Light curve obtained using aperture of enumeration and ResNet-18 with different variation of contamination source.}}
  \label{fig:apercomplc}
   \vspace{-0.9cm}
\end{figure}
}

\label{lastpage}

\bibliographystyle{raa}
\bibliography{bibtex}

\end{document}